\documentclass{ws-ijmpd}
\usepackage[super,compress]{cite}
\usepackage{amsmath}
\begin{document}


%
\catchline{}{}{}{}{}
%

\title{TRAVERSABLE AND STABLE WORMHOLE SOLUTIONS IN LOGARITHMIC $f(R)$ GRAVITY}

\author{Yusmantoro}

\address{Departmen of Physics, Universitas Gadjah Mada, Sekip Utara Bulaksumur\\
Yogyakarta, Indonesia\\
yusmantoro@mail.ugm.ac.id}

\author{Rian Julianto}

\address{Departmen of Physics, Universitas Gadjah Mada, Sekip Utara Bulaksumur\\
Yogyakarta, Indonesia\\
rianjulianto@mail.ugm.ac.id}

\maketitle


\begin{abstract}
This study aims to investigate the viability of a viable logarithmic $f(R)$ gravity for inflation in giving wormhole solutions. We consider a static and spherically symmetric spacetime and two different shape functions. We found that the model provides traversable wormhole solutions shown by its violation to  energy conditions. Violation to null energy condition indicates the existence of traversable wormhole and the existence of exotic matter. The equation of state parameter is less than -1 which shows the presence of phantom fluid. Furthermore, the stability of wormhole is analyzed using Tolman-Oppenheimer-Volkoff equation and we found that the wormhole solutions are stable. Finally, we conclude that the model is viable in giving wormhole solutions with the same parameter constraints as inflation. 
\end{abstract}




\section{Introduction}	
Wormhole is believed to be a hypothetical astrophysical object connecting two separated different areas in the universe. Wormhole is also known as Einstein-Rosen bridge which can be ragarded as a static solution of Einstein equations  \cite{shatski}. Geometrically, wormhole consists of a throat which connects two asymptotically flat spaces \cite{Novikov}. Although its existence remains unproven, wormhole is consistent with general relativity. Because, wormhole solution is provided by Einstein field equations.

Traversable wormhole solution was firstly proposed by Ref \refcite{morris} as a tool for teaching general relativity. It was an interesting tool for students because its property  allowing time travel.  According to their assumptions, wormhole was not only regarded as a pedagogical tool in teaching but also something having possibility to be built in the future by advanced civilization. They studied wormhole by using static spherically symmetric metric and assuming the existence of exotic matter. This leads to the violation to energy condition, especially is null energy condition. This is a physical consequence because general relativity does not give the traversable solutions without violating null energy condition \cite{Godani}.

Wormhole has become an interesting topic in cosmology and astrophysics due to its unique properties. The wormhole belongs to non-vacuum solutions obtained from Einstein Field Equations which is filled with exotic matter\cite{Samantha}.  Introducing a specific form of exotic matter is difficult problem in particle physics. Furthermore, it has not been found in experimental particle physics. However, the possibility of the existence of wormhole can still be analyzed in the framework of modified gravity due to its new degrees of freedom to general relativity. The existence of exotic matter is shown by the trace of energy momentum which violates null energy condition \cite{aziz}.

Modified gravity works on geometrical modification of gravity. Mathematically, one of the theory in modified gravity which can be regarded as the simplest theory is $f(R)$ gravity. In this theory, Ricci scalar in Einstein-Hilbert action is replaced by a function of $R$ namely $f(R)$ to give new degrees of freedom in order to describe cosmological phenomena which can’t be obtained using general relativity. Besides its simplicity, a lot of cosmological objects such as compact star \cite{zubair}, black hole \cite{hendi} , and neutron stars \cite{astashenok} can be studied very well using  $f(R)$ gravity.  We can conclude that $f(R)$ gravity is still becoming a good theoretical framework in cosmology and astrophysics.

There are numerous wormhole solutions obtained in the framework of $f(R)$ gravity. Ref \refcite{golchin} obtained asymptotically flat and hyperbolic wormhole solutions in some static $f(R)$ model using non-constant Ricci scalar $R$. Ref \refcite{capo}  studied wormhole solutions in power law $f(R)$ gravity with vanishing sound speed. Ref \refcite{kara}  derived an exact wormhole spacetime involving phantom scalar field in $f(R)$ gravity where the existence of tachyonic instability and ghost are avoided. Ref \refcite{shamir} investigated non-commutative wormhole solutions in $f(R)$ gravity providing the existence of dark matter and dark energy.

In this research, we are interested in studying the viability of a viable $f(R)$ model proposed by Ref \refcite{amin} in giving wormhole solutions. The model belongs to a class of logarithmic $f(R)$ gravity which successfully describes inflationary scenario. This research aims to investigate the viable wormhole solutions from this model by considering its inflation parameter constraints. Wormhole solutions will be obtained using two different shape functions and a constant redshift function. Furthermore, we will consider the static and spherically symmetric spacetime.

\section{Logarithmic $f(R)$ Gravity}
The action of modified $f(R)$ gravity in Planck unit is written as
\begin{equation}
	\begin{aligned}
	S&=\frac{1}{16\pi}\int \,d^{4}x f(R) +S_{m}\left(g_{\mu\nu},\psi_{m}\right)\\
	&=\int \,d^{4}x \sqrt{-g}\left[\frac{1}{16\pi}f(R) + \mathcal{L}_{m}\left(g_{\mu\nu},\psi_{m}\right) \right],
	\end{aligned}
\end{equation}
where $\mathcal{L}_{m}\left(g_{\mu\nu},\psi_{m}\right)$ is the Lagrangian density of matter and $\psi_{m}$ denotes the matter field which is minimally coupled to metric tensor $g_{\mu\nu}$.
Varying the action above with respect to the metric tensor, we obtain the field equation
\begin{equation}
	F(R)R_{\mu\nu}-\frac{1}{2}g_{\mu\nu}f(R)+\left[g_{\mu\nu}\Box-\nabla_{\mu}\nabla_{\nu}\right]F(R)=8\pi T_{\mu\nu},
\end{equation}
 $T_{\mu\nu}$  is the energy momentum tensor of anisotropic matter representing the matter source of wormholes. It can be written as
\begin{equation}
	T_{\mu\nu}=(\rho+p_{t})u_{\mu}u_{\nu}-p_{t}g_{\mu\nu}+(p_{r}-p_{t})V_{\mu}V_{\nu}
\end{equation}
$\rho$, $p_{r}$, and $p_t$ stand for energy density, radial pressure, and tangential pressure, respectively.
$V_{\mu}$  is the radial unit four vector whereas $u_{\mu}$ is the four velocity of the matter distribution. They have to satisfy the relation $u^{\mu}u_{\mu}=-1$ and $V^{\mu}V_{\mu}=1$.
Here, we will consider logarithmic $f(R)$ gravity taking the form as
\begin{equation}
	f(R)=R+\frac{m^{4}}{3M^{2}}\left[\frac{R}{m^{2}}-\ln\left(1+\frac{R}{m^{2}}\right)\right]
\end{equation}
where $m$ and $M$ are the free parameters. This model was taken from Ref \refcite{amin} which was initially used to explain inflation and gave the parameter constraints as $M\sim10^{-6}$ ,$10^{-4}\lesssim m \lesssim10^{-2}$ and $10^{2}\lesssim \frac{m}{M}  \lesssim10^{4}$. In this research, those constraints will be examined based on the viability of the model in giving plausible wormhole solutions.

\section{Static and Spherically Symmetric Wormhole Solutions in Logarithmic $f(R)$ Gravity}
The geometry of wormhole is defined by a static and spherically symmetric spacetime where the line element can be written as
\begin{equation}
	\centering
	ds^2=-e^{2\Phi(r)}dt^2+\frac{dr^2}{1-\frac{b(r)}{r}}+r^2\left(d\theta^2+\sin^2\theta d\phi^2\right). 	
\end{equation}
$\Phi(r)$ denotes the redshift function determining the gravitational redshift and $b(r)$ is the shape function which describes the spatial shape of wormhole. For the wormhole geometry, Ricci scalar $R$ and shape function is related by 
\begin{equation}
	R=\frac{2b'(r)}{r}.
\end{equation}
The shape function has to satisfy the conditions below
\begin{itemize}
\raggedright
\setlength\itemsep{1em}
\item $b(r_{0})=r_{0}$
\item $\frac{b(r)-b'(r)r}{b(r)^{2}}>0$
\item $b'(r_{0})-1\leq0$
\item $\lim_{r \to \infty} \frac{b(r)}{r}=0$
\end{itemize}
$b'(r)$ stands for $\frac{db(r)}{dr}$ and $r_{0}$ is the radius of the wormhole throat.
\subsection{The Energy Conditions}
The existence of traversable wormhole is shown by its violation to energy conditions. 
Violation to null energy condition leads to traversability of wormholes as the negative values of (NEC) prevents the throat of wormholes to shrink \cite{Yousaf}. Not only does wormhole geometry violate null energy condition but also weak energy condition (WEC)\cite{lobo}. The combinations of energy density, radial pressure, and tangential pressure provides strong energy conditions. Therefore, in this paper, we will consider three energy conditions namely null energy condition (NEC), weak energy condition (WEC), and strong energy condition (SEC) written as
\begin{itemize}
	\raggedright
	\setlength\itemsep{1em}
	\item $\rho+p_{r}\geq0, \rho+p_{t}\geq0$ (NEC)
	\item $\rho\geq0, \rho+p_{r}\geq0, \rho+p_{t}\geq0$ (WEC)
	\item $\rho+p_{r}\geq0, \rho+p_{t}, \rho+p_{r}+2p_{t} \geq0$ (SEC).
\end{itemize}
To obtain them, we need to formulate the energy density $\rho$, radial pressure $p_r$, and tangential pressure derived from field equations.  They can be obtained using the following formulas \cite{sam}
\begin{equation}
	\begin{aligned}
&\rho(r)=\frac{df}{dR}\frac{b'(r)}{r^{2}}\\
&p_{r}=-\frac{df}{dR}\frac{b(r)}{r^{3}}
+\frac{d^{2}f}{dR^{2}}\frac{rb'(r)-b(r)}{2r^{2}}-\frac{d^{3}f}{dR^{3}}\left(1-\frac{b(r)}{r}\right)\\
&p_{t}=-\frac{1}{r}\frac{d^{2}f}{dR^{2}}
\left(1-\frac{b(r)}{r}\right)+\frac{1}{2r^{3}}\frac{df}{dR}\left(b(r)-rb'(r)\right).		
	\end{aligned}
\end{equation}
Here, we will use two shape functions i.e.
$b(r)=\frac{r_{0}\ln(r+1)}{\ln(r_{0}+1)}$ proposed by Ref \refcite{Samantha} and $b(r)=r_{0}\left(\frac{r}{r_{0}}\right)^{\frac{1}{2}}$ taken from Ref \refcite{moraes}. Meanwhile, the redshift function is taken to be a constant for both cases.

\subsubsection{Case I}
We will consider firstly the logarithmic shape function $b(r)=\frac{r_{0}\ln(r+1)}{\ln(r_{0}+1)}$. The energy density, radial pressure, and tangential pressure are determined by equation (7).
\begin{equation}
\rho=\frac{\left(M^{2}m^{2}r^{2}(r+1)\ln(r_{0}+1)+\left(M^{2}+\frac{m^{2}}{3}\right)r_{0}\right)r_{0}}{{r^{2}(r+1)\left(m^{2}(r+1)\ln(r_{0}+1)r^{2}+2r_{0}\right)M^{2}\ln(r_{0}+1)}}
\end{equation}
\begin{equation}
\begin{aligned}
p_{r}&=-\frac{1}{r^{3}(r+1)\left(m^{2}(r+1)\ln(r_{0}+1)r^{2}+2r_{0}\right)^{3}M^{2}\ln(r_{0}+1)}\\
\times &\left(r_{0}\left(A+6(B)r_{0}r^{3}(r+1)m^{4}\ln(r_{0}+1)
+8\left(M^{2}+\frac{m^{2}}{3}\right)r_{0}^{3}(r+1)\ln(r+1)\right)\right),
\end{aligned}
\end{equation}
where $A$, $B$, and $C$ are
\begin{equation}
	\begin{aligned}
		\nonumber
	A=&r^{5}\left(M^{2}r(r+1)^{3}\ln(r+1)+8r^{2}+\frac{32r}{3}+4\right)(r+1)m^{6}\ln(r_{0}+1)^{3}\\
	B=&\left(\left(\left(M^{2}-\frac{25m^{2}}{18}\right)r^{2}+\left(2M^{2}-\frac{11m^{2}}{6}\right)r+M^{2}-\frac{2m^{2}}{3}\right)r\ln(r+1)+\frac{m^{2}r^{3}}{6}
	\right.\\
	&\left.
	+\frac{m^{2}r^{2}}{9}-\frac{4r}{3}-\frac{4}{9}\right)\\
	C=&\left(\left(\left(M^{2}+\frac{13m^{2}}{18}\right)r+M^{2}+\frac{m^{2}}{3}\right)(r+1)\ln(r+1)
	\right.\\
	&\left.
	+\frac{\left(r+\frac{2}{3}\right)rm^{2}}{6}\right)
	\end{aligned}
\end{equation}
\begin{equation}
\begin{aligned}
	p_{t}=\frac{\left(r_{0}\left(E+4F+4G\right)\right)}{2r^{3}(r+1)\left(m^{2}(r+1)\ln(r_{0}+1)r^{2}+2r_{0}\right)^{2}M^{2}\ln(r_{0}+1)}
	\end{aligned}
\end{equation}
where $E$,$F$, and $G$ are
\begin{equation}
	\nonumber
	\begin{aligned}
		E=&\left((r+1)^{2}M^{2}r\ln(r+1)-M^{2}r^{3}-M^{2}r^{2}+4r+\frac{8}{3}\right)\\
		F=&\left(\left(\left(M^{2}-\frac{5m^{2}}{6}\right)r+M^{2}-\frac{m^{2}}{2}\right)
		\ln(r+1)-\left(M^{2}+\frac{m^{2}}{6}\right)r\right)\\
		&r_{0}r^{2}(r+1)m^{2}\ln(r_{0}+1)\\
		G=&\left(M^{2}+\frac{m^{2}}{3}\right)\left((r+1)\ln(r+1)-r\right)r_{0}^{2}.
	\end{aligned}
\end{equation}
 
\subsubsection{Case II}
For the second case, we use the shape function $b(r)=r_{0}\left(\frac{r}{r_{0}}\right)^{\frac{1}{2}}$
\begin{equation}
\rho=\frac{3M^{2}\sqrt{\frac{r}{r_{0}}}m^{2}r^{2}+3M^{2}+m^{2}}{6\sqrt{\frac{r}{r_{0}}}r^{2}M^{2}
\left(m^{2}\sqrt{\frac{r}{r_{0}}}r^{2}+1\right)
}
\end{equation}
\begin{equation}
	\begin{aligned}
	p_{r}=\frac{1}{24r^{2}r_{0}\sqrt{\frac{r}{r_{0}}}M^{2}\left(m^{2}\sqrt{\frac{r}{r_{0}}}r^{2}+1\right)^{3}}\left(-24M^{2}r^{7}\sqrt{\frac{r}{r_{0}}}m^{6}-70\sqrt{\frac{r}{r_{0}}}m^{6}r^{5}
	\right.\\
	\left.
	-72M^{2}m^{4}r^{5}+67m^{6}r^{5}-72M^{2}r^{2}r_{0}\sqrt{\frac{r}{r_{0}}}m^{2}-41\sqrt{\frac{r}{r_{0}}}r^{2}m^{4}r^{3}-24M^{2}r_{0}-8m^{2}r_{0}\right)
	\end{aligned}
\end{equation}
\begin{equation}
	\begin{aligned}
		p_{t}=\frac{1}{12r^{2}\sqrt{\frac{r}{r_{0}}}M^{2}\left(m^{2}\sqrt{\frac{r}{r_{0}}}r^{2}+1\right)^{2}r_{0}}
		\left(3M^{2}m^{4}r^{5}+6M^{2}r^{2}r_{0}\sqrt{\frac{r}{r_{0}}}m^{2}
		\right.\\
		\left.
		-9\sqrt{\frac{r}{r_{0}}}r^{2}m^{4}r_{0}
		+10m^{4}r^{3}+3M^{2}r_{0}+m^{2}r_{0}\right)
	\end{aligned}
\end{equation}	
\newpage
\subsection{Graphs Representation of Energy Conditions}
Combining Eq $8$ to $13$ we are able to fomulate the energy conditions. The behavior of energy conditions can be studied based on their plots presented in Fig 1 for case I and Fig 2 for case II, respectivelty.

\begin{figure}[h]
	\centering
	\includegraphics[scale=1.3]{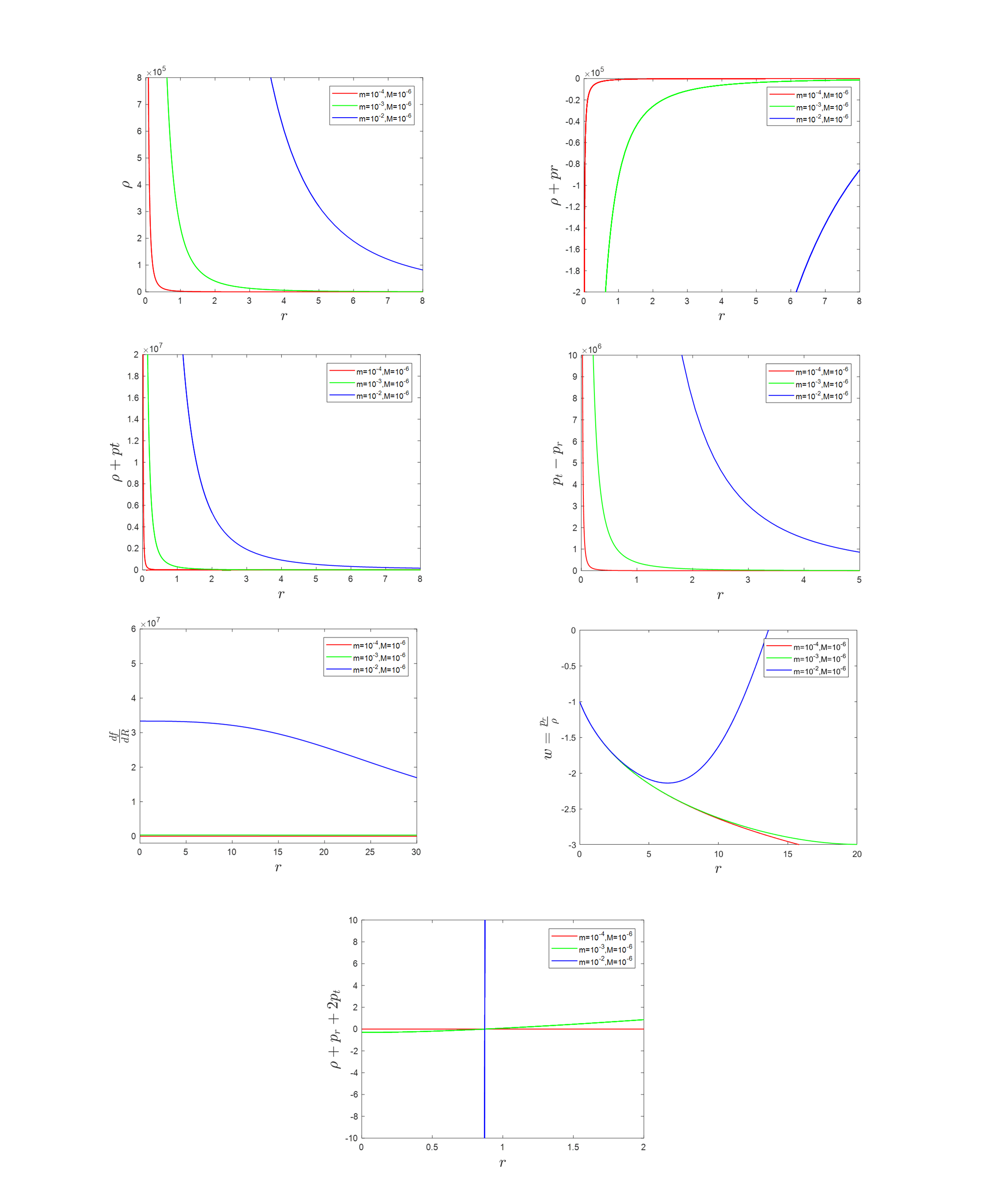}
	\caption{Case I : Plots for energy density $(\rho)$, radial null energy condition $(\rho+p_{r})$, tangential null energy condition $(\rho+p_{t})$ , anisotropy parameter $(\Delta=p_{t}-p_{r})$, $(\frac{df}{dR})$, equation of state parameter $(w)$, and strong energy condition $(\rho+p_{r}+2p_{t})$.}
	\label{fig:x cubed graph}
\end{figure}

\newpage
\begin{figure}[h]
	\centering
	\includegraphics[scale=1.3]{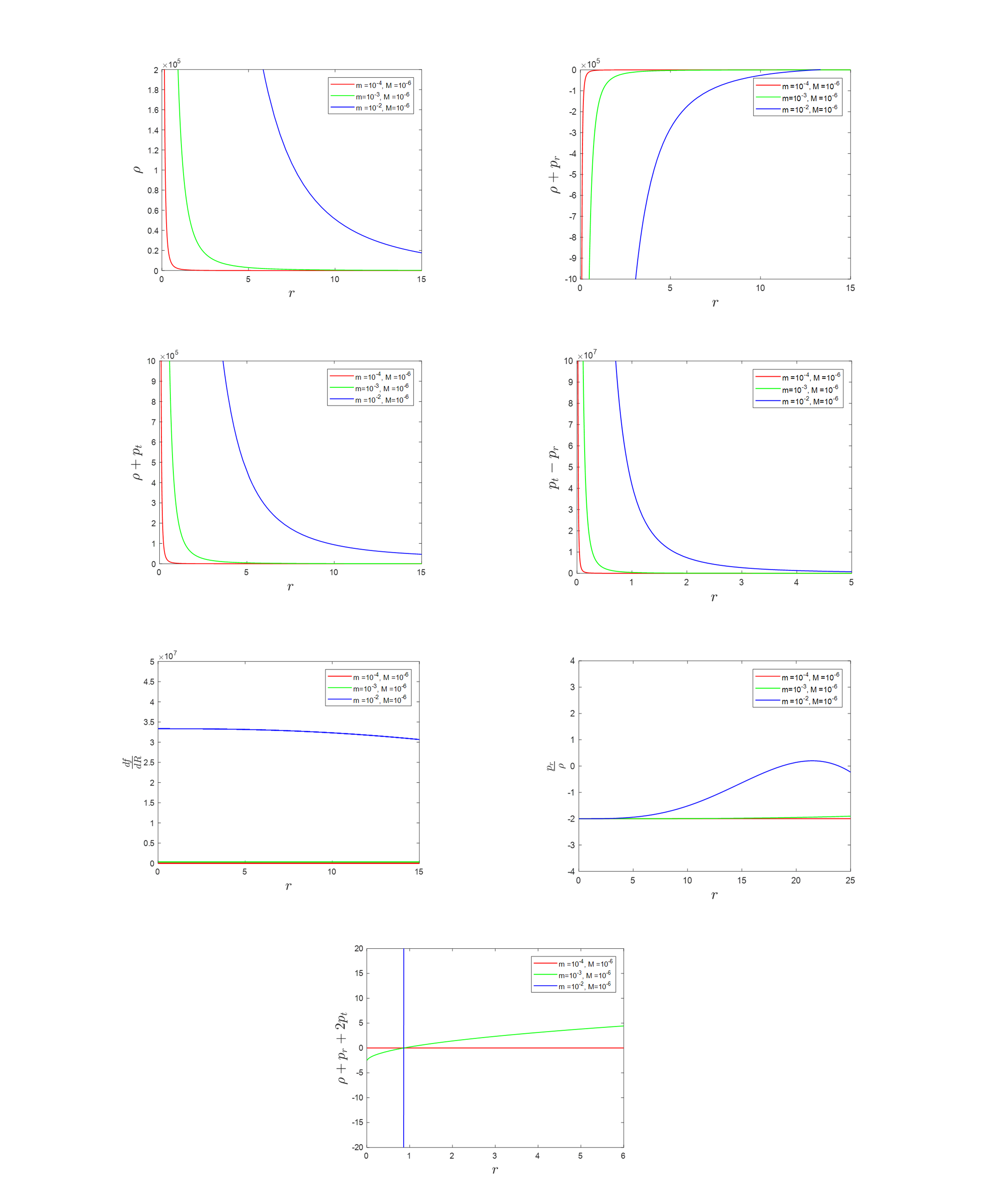}
	\caption{Case II : Plots for energy density $(\rho)$, radial null energy condition $(\rho+p_{r})$, tangential null energy condition $(\rho+p_{t})$ , anisotropy parameter $(\Delta=p_{t}-p_{r})$, $(\frac{df}{dR})$, equation of state parameter $(w)$, and strong energy condition $(\rho+p_{r}+2p_{t})$.}
	\label{fig:x cubed graph}
\end{figure}

\section{Stability of wormhole solutions}
To investigate the stability of wormhole solutions, we may use Tolman-Oppenheimer-Volkov (TOV) equation \cite{Mustafa, Hassan, Shweta,Rahaman, Banerjee}. It takes the following form
\begin{equation}
	\frac{dp_{r}}{dr}+\frac{\Pi'}{2}\left(\rho+p_{r}\right)+\frac{2}{r}\left(p_{r}-p_{t}\right)=0,
\end{equation}
where $\Pi'=2\Pi$. Three terms of TOV equation stand for gravitational force, hydrostatic force, and anisotropic force, respectively. Those are written as 
\begin{equation}
	F_{grav}=-\frac{\Pi'}{2}\left(\rho+p_{r}\right) ;\hspace{4 mm} F_{hyd}=-\frac{dp_{r}}{dr} ;\hspace{4 mm} F_{an}=\frac{2}{r}\left(p_{t}-p_{r}\right).
\end{equation}
In the case of constant redshift function, gravitational force will be zero. Consequently, TOV equation can be simplified as
\begin{equation}
	\begin{aligned}
		\frac{dp_{r}}{dr}+\frac{2}{r}\left(p_{r}-p_{t}\right)&=0\\
		F_{hyd}+F_{an}&=0
	\end{aligned}
\end{equation}
which means that hydrostatic and anisotropic force have to cancel each other. Their behavior can be analyzed in the two following graphs.
\begin{figure}[h]
	\centering
	\includegraphics[scale=0.8]{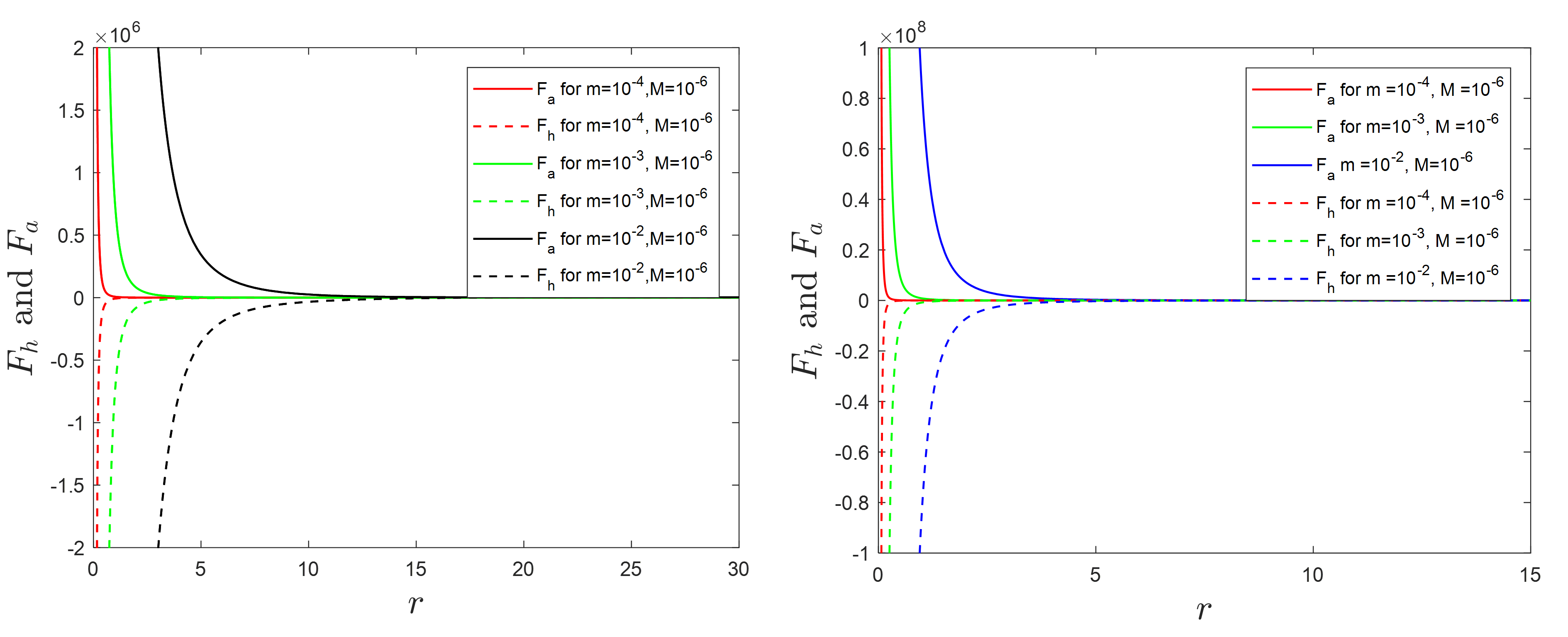}
	\caption{Plots for anisotropic force $F_{a}$ (dashed line) and hydrostatic force $F_{h}$ (solid line) for Case I (left) and Case II (right).}
	\label{fig:x cubed graph}
\end{figure}

\section{Result and Discussion}
Both shape functions represent the similiar graphs of energy conditions. They show the violation to radial null energy condition $(\rho+p_{r})\leq0$ which indicates the existence of exotic matter. Tangential null energy condition  is satisfied due to $\rho+p_{t}\geq0$. The violation to radial null energy condition leads to the vioation to weak energy condition and the strong energy condition. Furthermore, we can see that $\rho+p_{r}+2p_{t}\leq0$ for $r<1$. 
In both cases, we know that $\frac{df}{dR}>0$ . Therefore, the non-existence theorem is satisfied. The anisotropy parameter $\left(\Delta\right)$ is always positive and descreasing for both case. 
It can be concluded that the existence of wormhole is not affected by anisotropy \cite{shamir}. The equation of state parameter $(w)$ is always less than -1 indicating that wormhole is filled with phantom fluid.

In term of stability, it can be seen that graphically in Fig 3. The hydrostatic and anisotropic force have opposite value at each points indicating that their resultant is vanish everywhere. These force balance each other. Therefore, we can conclude that from both shape functions wormhole solutions are stable.
\section{Conclusion}
We have analyzed the viability of a viable logarithmic $f(R)$ model for inflation in giving wormhole solutions using static and spherically symmetric spacetime and two shape functions. The parameter constraints in inflation of the model are considered here. We obtain that from two shape functions, the model provides traversable wormhole solutions shown by its violation to energy conditions.   The violation to null energy condition leads to the existence of exotic matter. The value of equation of state parameter $(w)$ confirms the presence of phantom fluid. Furthermore, using Tolman-Oppenheimer-Volkov (TOV), hydrostatic force and anisotropic force cancel each other. This condition leads to the stable condition of wormhole solutions. We can conclude that besides its viablity to explain inflation, logarithmic $f(R)$ gravity is a viable model for giving traversable and stable wormhole solutions.

\end{document}